\documentclass[9pt,twocolumn,twoside]{pnas-new}

\templatetype{pnasresearcharticle} 

\usepackage{caption}
\usepackage{subcaption}
\usepackage{graphicx}
\usepackage{comment}

\title{Academic Co-authorship is a Risky Game}

\author[a,1]{Teddy Lazebnik}
\author[a]{Stephan, Beck}
\author[b]{Labib Shami} 

\affil[a]{Department of Cancer Biology, Cancer Institute, University College London, London, UK}
\affil[b]{Department of Economics, Western Galilee College, Acre, Israel}

\leadauthor{Lazebnik} 

\significancestatement{Publication authorship is one of the main recognitions researchers obtain for their research. It is therefore paramount to have a robust and acceptable system in place that ensures fairness and due process for disputes. The current system largely relies on the integrity of researchers but often fails when disputes arise. The use of an unbiased, data-driven model could overcome this limitation. In this study, we developed the first game-theory-inspired model to explore the dynamics of co-authorship disputes. The results are alarming and suggest that the current academic practices are not fit for purpose. Worryingly, our model shows that the number of potential disputes increases as the number of authors participating in the research increases, which is the global trend.}

\authorcontributions{T. L.: Conceptualization, Design, Methodology, Software, Visualization, Formal analysis, Investigation, Writing- Original draft preparation. S. B.: Project administration, Writing- Reviewing. L. S.: Methodology, Formal analysis, Writing- Original draft preparation. All authors reviewed the manuscript and approved the final version.}
\authordeclaration{The authors declare no competing interests.}
\correspondingauthor{\textsuperscript{1}To whom correspondence should be addressed. E-mail: t.lazebnik@ucl.ac.uk}


\begin{abstract}
Conducting a research project with multiple participants is a complex task that involves not only scientific but also multiple social, political, and psychological interactions.  This complexity becomes particularly evident when it comes to navigating the selection process for the number and order of co-authors on the resulting manuscript for publication due to the current form of collaboration dynamics common in academia. There is currently no computational model to generate a data-driven suggestion that could be used as a baseline for understating these dynamics. To address this limitation, we have developed a first game-theory-based model to generate such a baseline for co-authorship. In our model, co-authors can issued an ultimatum to pause the publication of the manuscript until the underlying issue has been resolved. We modeled the effect of issuing one or more ultimatums and showed that they have a major impact on the ultimate number and position of co-authors and the length of the publication process. In addition, we explored the effect of two common relationships (student-advisor and colleague-colleague) on co-authorship scenarios. The results of our model are alarming and suggest that the current academic practices are not fit for purpose. Where they work, they work because of the integrity of researchers and not by a systematic design. 
\end{abstract}

\begin{document}

\maketitle
\ifthenelse{\boolean{shortarticle}}{\ifthenelse{\boolean{singlecolumn}}{\abscontentformatted}{\abscontent}}{}

\dropcap{S}cientific collaboration is one of the pillars of a productive research environment \cite{lee2005impact} and co-authoring is considered one of the most tangible forms of scientific collaboration \cite{dehdarirad2017research,de2021measure,spreaded}. Over the past two decades, the number of team-authored manuscripts has increased, consistently outnumbering solo author studies \cite{wuchty2007increasing,amjad2017standing}. Several factors account for this increase in multi-author academic manuscripts including but not limited to more cross-disciplinary research and increasing demands on faculty members within academia to publish \cite{von2017academe}. Research collaborations rely on the assumption that the greater the number of authors working together on a multi-disciplinary research question, the greater the chances of obtaining better and more competitive results. This assumption intensifies as the complexity of the research and the level of competitiveness among scholars increase \cite{acedo2006co,zhang2018understanding}.

Various studies have investigated the sociological motives behind increase in co-authorship \cite{li2018important,zhang2018understanding,cugmas2020scientific,wieczorek2021better}. Many of these studies have been devoted to the challenges and problems arising from collaborations between researchers. These studies have revealed the “dark side” of scientific collaborations, They have highlighted various challenges ranging from ethical and practical issues to the abuse of subordinates and the use of academic status to advance one's self-interests, exposing several shortcomings of the widely accepted authorship system \cite{shrum2001trust,bozeman2004scientists,bozeman2012dark,youtie2014social}. \cite{borry2006author} show that as the number of authors involved increases, the magnitude of these challenges increases proportionally. Therefore, the growing number of multi-author studies is undermining the effectiveness of the long-standing scientific authorship system.

Tackling these challenges requires identifying the motivations and dynamics involved in any academic co-authorship project. One method of doing so is using game theory. Simply put, each co-authorship project involves the investment of time, efforts, funds, and expertise from the authors in return for the expected benefit of publishing the manuscript \cite{lawrence2002rank}. 

During the writing of the study, disagreements may arise regarding the order of the authors on the authors' list, which reflects the amount of credit and thus the utility each author receives from publishing the manuscript. In a commentary, \cite{avila2014bullying} argues that sometimes senior researchers engage in unethical behavior, \textit{bullying} their younger fellows and coercing them into changing the order of the authors on the final manuscript \cite{je2021sidelined, street2010credit}. In many cases, these unethical behaviors and conflicts may obstruct the publication process \cite{primack2014editorial}.
To the best of our knowledge, despite the extensive literature that discusses research collaborations, no study has addressed this aspect in detail. In addition, no explanations have been suggested that would help us to understand the dynamics of the disagreement. Finally, no model has been proposed that would determine the boundaries of the problem so we can understand its causes and shed light on an issue that has hitherto enjoyed a certain ambiguity. Indeed, scientometrics studies have used cooperative game theory methods, especially the \textit{Shapley value}, in solving problems of publication credit allocation \cite{tol2012shapley, papapetrou2011shapley, karpov2014equal, gauffriau2021counting}. These studies, however, did not assume disagreement among the authors but suggested a division of the publication credit that could be agreed upon by all.

Hence, in this study, we propose the first game-theory-based model for disputes of co-authorship credit that uses concepts from the ultimatum game model \cite{espin2015short,murnighan2008fairness,henrich2006costly}. This experimental set-up is commonly used to explore a wide range of human bargaining behavior involving two or more individuals, where one makes a proposal and the other responds. Most often, the game revolves around splitting a resource between the one individual who proposes a division of the resource between himself/herself and others who must to decide whether to accept or reject the offer. In most cases, the game ends with neither party receiving any resources. Given our research topic, such a game structure has the potential to enrich the environment in which the discussion about determining the order of the authors takes place. Thus, in academic co-authorship settings, such as ours, an ultimatum can be defined as one of the author's demands to upgrade his/her position on the authors' list. If the other authors do not agree to the ultimatum, he/she would oppose the submission of the research to an academic journal, resulting in wasted time and resources for all participating authors.

\section*{Results}
We developed a game-theory-based model for the case in which an author who is a co-author in an academic project would issue an ultimatum (IAU) to improve his/her position on the authors' list and obtain the desired result. The proposed model assumes all the authors are rational and fully aware of the state of the other authors. Each author contributes to the completion of the project (i.e., submitting the manuscript) and is rewarded according to his/her position on the authors' list. For ease of exposition, a formal definition of the model and its implementation as an computer simulation are presented in the Materials and Methods section. 

\subsection{The number of authors influences the likelihood of issuing an ultimatum}
\begin{figure*}[!ht]
    \centering
    \includegraphics[width=0.99\textwidth]{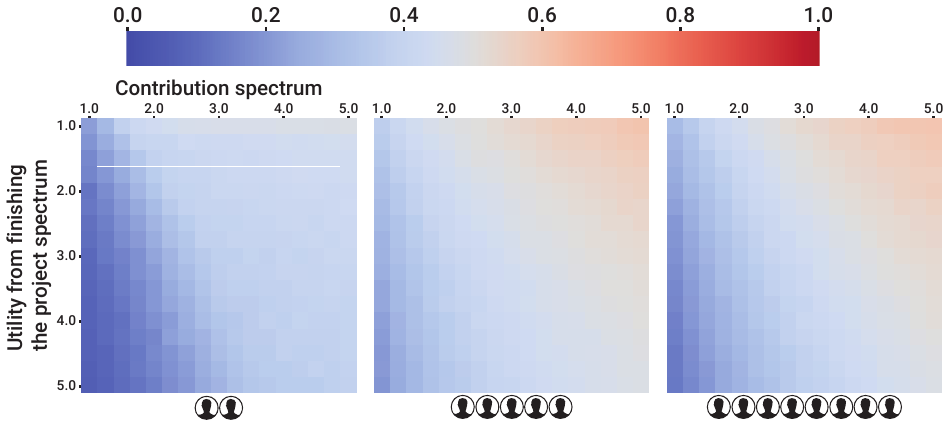}
    \caption{The influence of the authors' contributions and utilities from finishing the project on the rate that at least one author would raise an ultimatum, divided into (a) two, (b) five, and (c) eight authors (from left to right). The results are shown as an average of \(n=10,000\) simulations.}
    \label{fig:1}
\end{figure*}

Figure \ref{fig:1} illustrates the results of the simulations of our model indicating the rate at which at least one author would IAU as a function of the authors' contribution and utility from finishing the project. Each value is an average of \(n=10,000\) simulations. There are two factors depicted in the figure: the utility spectrum, meaning the spread between the author who benefits the most from finishing the project and the author who benefits the least, and the contribution spectrum, meaning the spread between the author who contributes the most to the project and the author who contributes the least. As the figure indicates, when the utility spectrum declines and the contribution spectrum increases, the probability that at least one author will IAU increases. Stated mathematically, the formula \(P = 0.22 -0.04U + 0.09*C\) (\(R^2 = 0.85)\) where \(P\) is the average rate of IAU, \(U\) is the utility spectrum, and \(C\) is the contribution spectrum, for the case of five authors. Moreover, as the number of authors increases the average rate of IAU increases logarithmically. Thus, \(P = 0.18ln(A) + 0.12\) (\(R^2 = 0.78\)) where \(A\) is the number of authors.

\subsection{The sensitivity of the project's parameter}
\begin{figure*}[!ht]
    \centering
    \includegraphics[width=0.99\textwidth]{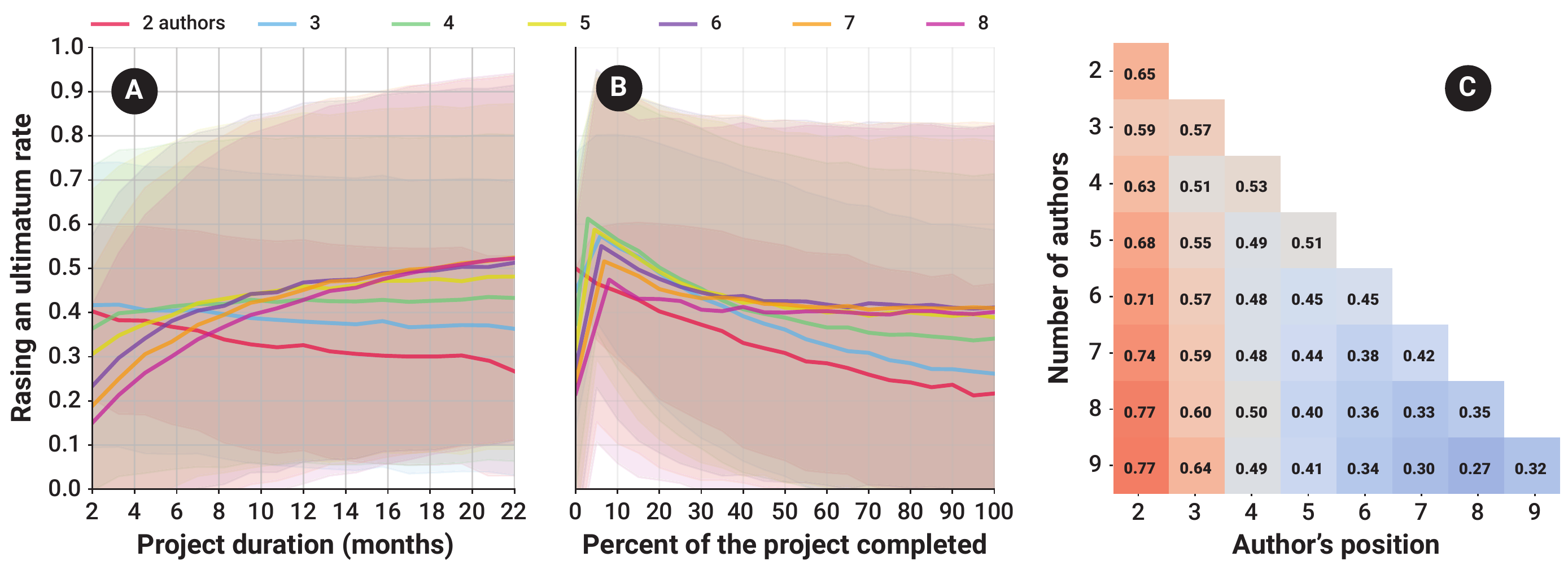}
    \caption{The influence of the (a) project's expected duration, (b) project's completion state, and (c) author's position on the rate it is beneficial to raise an ultimatum for at least one author, divided into the number of authors in the project. The results are shown as mean \(\pm\) standard deviation of \(n=100,000\) reparations.}
    \label{fig:2}
\end{figure*}

The results in Figure \ref{fig:1} indicates the average of a wide range of possible co-authorship configurations such as current point in the project, its duration, and the position of the ultimatum issuing author on the authors' list. This last factor plays a key role in the rate at which an author will IAU, as shown in Figure \ref{fig:2}, where the results are shown as the mean \(\pm\) standard deviation of \(n=100,000\) simulations. Figure \ref{fig:2}A shows that for projects with two and three authors, as the project's duration lengthens, the average rate of IAU declines linearly. Nevertheless, for the case of four or more authors, the trend mirror and the rate increase logarithmically relative to the project's duration. In a similar manner, Figure \ref{fig:2}B indicates that the state of the project relative to its completion influences the IAU rate significantly. For projects with two authors, as the project progresses, the average rate of raising an ultimatum is decreasing logarithmically. However, for three or more authors, there is a sharp increase at the beginning of the project up to the point where about 10\% of the project has been completed, followed by a logarithmic decrease. Figure \ref{fig:2}C indicates that an author will IAU as a function of the number of authors and his/her position on the authors' list such that each row in the matrix is obtained by averaging the results of \(n=100,000\) simulations. Note that as the number of authors increases, the IAU rate of previous positions increases. However, on average, each new position has a lower rate than the previous ones. 

\subsection{Common co-authorship cases}
Two common types of co-authorship research projects are between a student and his/her academic advisor(s) and between two groups of colleagues from different disciplines \cite{lawrence2002rank}. The first is a by-product of the academic training of new researchers while the latter is rooted in the desire for more multi-disciplinary projects. While the interactions and relationships in both cases are complex and influenced by a large number of factors \cite{weijer2003unethical,acedo2006co,zhang2018understanding}, we focused on four cases defined by two parameters: either the student contributes to the research project in equal measure as the advisor or significantly more, and either the student's utility from publishing the manuscript is similar to that of the advisor or significantly greater. We denoted these cases as \(SA_1, SA_2, SA_3\), and \(SA_4\), respectively. These cases are also tests for the case of a student with two advisors, denoted as \(SA_5, SA_6, SA_7\), and \(SA_8\). In addition, we extended these cases for two groups of two authors denoted as \(P_1, P_2, P_3\), and \(P_4\). Table \ref{table:cases} summarizes these cases and their parameters.

\begin{table}[!ht]
\centering
\begin{tabular}{|p{0.1\textwidth}|p{0.12\textwidth}|p{0.07\textwidth}|p{0.06\textwidth}|}
\hline
\textbf{Case} & \textbf{Contribution spectrum} & \textbf{Utility spectrum} \\ \hline
\(SA_1, \; SA_5, \; P_1\) & Similar - [1, 1.5] & Similar\\ \hline
\(SA_2, \; SA_6, \; P_2\) & Different - [1.5, 3] & Similar  \\ \hline
\(SA_3, \; SA_7, \; P_3\) & Similar & Different  \\ \hline
\(SA_4, \; SA_8, \; P_4\) & Different & Different \\ \hline
\end{tabular}
\caption{The special cases and their parameters.}
\label{table:cases}
\end{table}

\begin{figure*}[!ht]
    \centering
    \includegraphics[width=0.6\textwidth]{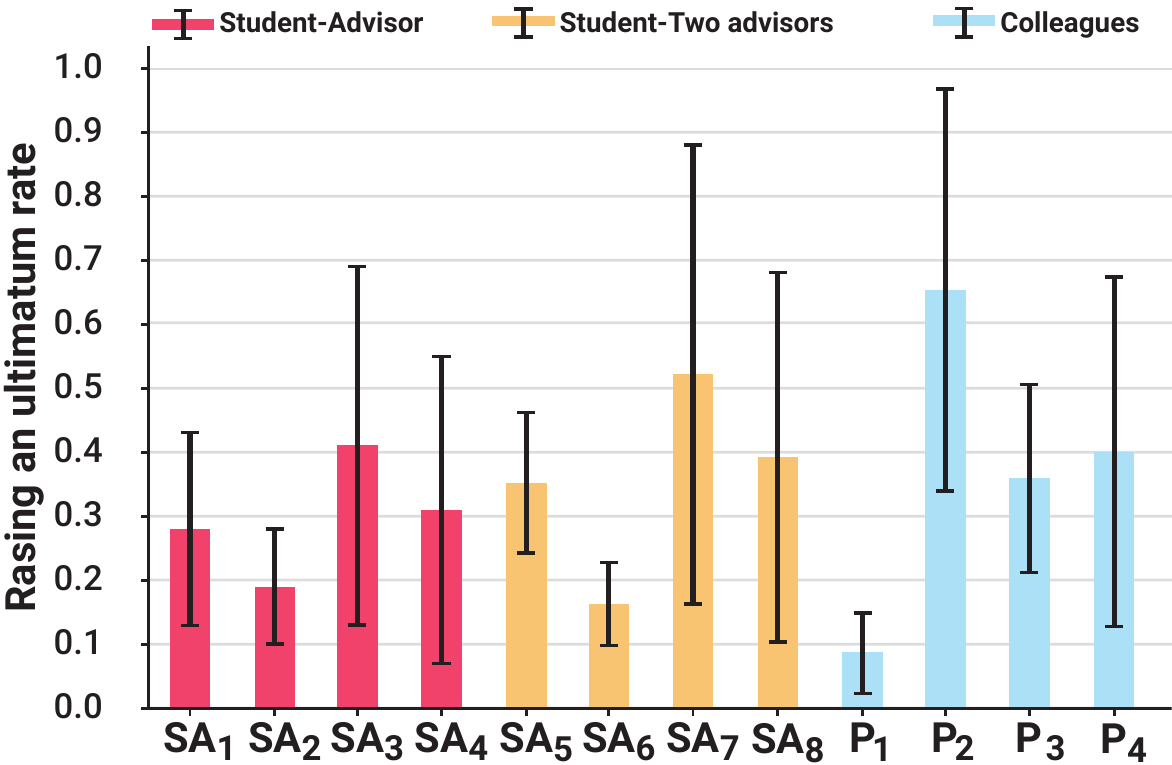}
    \caption{The rate an author would raise an ultimatum for different student-advisor(s) and colleagues co-authorship cases. The results are shown as average \(\pm\) standard deviation of \(n=100,000\) simulations.}
    \label{fig:3}
\end{figure*}

Figure \ref{fig:3} illustrates the rate at which one of the authors would IAU for each these cases, where the results are shown as an average \(\pm\) standard deviation of \(n=100,000\) simulations. Comparing the cases of a student with a single advisor (\(\{SA_i\}_{i=1}^4\)) and a student with two advisors (\(\{SA_i\}_{i=5}^8\)) shows they are pair-wise different in a statistically significant manner based on a two-tailed paired T-test, \(p < 0.005\). Moreover, equal authors who contribute similarly and obtain a similar utility from publishing the manuscript (i.e., \(P_1\)) results in an average IAU rate of around 8\%. This result underscores that even the most straightforward case is not free of possible IAUs.  

\section*{Discussion}
Taken together, the results in Figures \ref{fig:1}-\ref{fig:3} suggest that the current form of co-authorship in which the consent of all authors at any given point in the project is necessary to achieve the shared objective of finishing the project and submitting the manuscript has an enormous number of configurations in which one author can IAU and succeed in doing so. Although on average the system's dynamics encourage authors to IAU due to the complexity of co-authorship and a large number of parameters, the differences, as reflected in the standard deviations in Figures \ref{fig:2} and \ref{fig:3}, indicates that co-authorship all over the IAU range is statistically feasible and even quite common. 

In order to clarify, let us consider a research group consisting of four authors working on a study expected to last 22 months. All of the authors have a similar or identical utility from finishing the project. Moreover, the study is in its early stages when less than 10\% of it has already been completed, and one of the authors has invested five times more than the last-ranked author. The middle graph in the top right-hand corner of Figure \ref{fig:1}, combined with Figures \ref{fig:2}A and \ref{fig:2}B (yellow line) depict this situation. This example represents a research group most likely to have at least one of its members, especially the author who contributed the least, IAU. It is not difficult to find such situations in reality. Any study that requires the expertise of an individual whose role begins only after the data have been collected (a condition that describes the progress of up to 10\% of the study) falls into this category \cite{lawrence2002rank}. Hence, according to our analysis, this researcher may demand an improvement in his/her position by issuing an ultimatum, achieveing his/her goal. Thus, given that some authors contribute more than others, for more or less the same expected utility, the authors who contributed less are by design encouraged to IAU. They have little to lose. Moreover, these who contributed more would agree to the ultimatum to save at least some of the hard work they have invested and not leave the project empty-handed. 

In general,  as Figure \ref{fig:1} indicates, the rate of IAU increases as the number of authors increases. However, as Figure \ref{fig:3}C reveals, the rate of IAU does not increase identically for all of the authors. It depends on their position on the authors’ list. These results agree with the overall trend that having more authors usually makes writing the paper more difficult \cite{many_authors_issue}. Moreover, as Figure \ref{fig:2}A illustrates, the greater the number of authors and the longer the project's duration, the higher the IAU rate. This outcome underscores the complexity of managing the dynamics of large research groups. Such groups tend to work on large-scale and therefore usually longer projects that result in a two-fold increase in the IAU rate. 

When exploring the student-advisor(s) cases, the dominant configuration is one in which the student has a much greater utility compared to the advisor(s). Typically, the student is writing a paper for his/her master's or doctorate studies. In contrast, for the advisor(s) this is just another manuscript. In such cases, the student may conduct the study while the advisor handles the pedagogic, financial, and administrative work (i.e., cases \(SA_3\) and \(SA_7\)). In these situations, as Figure \ref{fig:3} illustrates, there is 41\% and 51\% IAU rate. Another scenario is one in which relatively academically younger advisor(s) has make a contribution similar to that of their students and have a similar utility (cases \(SA_2\) and \(SA_6\)). These advisors have a statistically significant (\(p < 0.005\)) lower IAU rate compared to their more academically mature colleagues. Nevertheless, their IAU rate is still high at 28\% and 34\% on average for one and two advisors, respectively. Unsurprisingly, for the case of four colleagues who made similar contributions and have a similar utility (case \(P_1\)), the IAU rate is as low as 8\%, while a large contribution spectrum (case \(P_2\)) results in a 65\% IAU rate on average. 

Moreover, the results explain why academically young researchers might agree to include a more senior researcher in their study, even placing him/her at the top of the authors' list, although his/her contribution to the project might be relatively small. The students' tacit consent might be given to ensure that the supervisor to appear as the first author, although writing the research is primarily the product of their thought, becomes rational given the fact that the supervisor's position not at the top of the list may provoke his/her future opposition to finishing the project and submitting the manuscript to a journal. Hence, their consent makes it more likely that the research will be published in due course.

One way to tackle the shortcoming in the current model of academic co-authorship is to introduce an outside influence to the system that regulates indications about the authors' contributions and their positions on the authors' list. One example is signing a contract at the beginning of the project that clearly states the contribution of each of the authors and his/her position on the authors' list accordingly \cite{primack2014editorial}. Thus, even if one of the authors attempts to engage in some exploitation during the study, the contract prevents him/her from succeeding. That said, in practice, this solution is not a silver bullet and suffers from a large number of issues \cite{law_1,law_2}. First, academic and research institutions should develop guidelines for these contracts, obviating the possibility that an uneven distribution of power might dictate the terms. Second, trying to enforce such contracts involving authors from different institutes and countries is not straightforward. Third, research projects are by nature unpredictable, adding more legal and technical complexity to the proposed remedy. Nonetheless, despite these challenges, the introduction of contracts to new disciplines has shown a significant reduction in disputes in the past \cite{law_good_1,law_good_2}. Another remedy that is gaining popularity is the requirement of journals to specify the nature of each author's contribution, which sheds some light on how credit should be distributed \cite{credit}. A future work might explore these and other remedies to this challenge with regard to their effectiveness and feasibility.

In this study, we assume each author is aware of the accurate state of all of the other authors in the project. This assumption is not unreasonable as each author is able to assess these values using prior knowledge of the academic system and background knowledge about the other authors to some degree. Nevertheless, in reality, there are holes in this assumption. For example, the duration of the project can only be estimated. Similarly, one author might assess the contributions of the other authors differently than the author assesses them himself/herself. This situation results in a partially observable and noisy state for all authors in the project. As such, an attempt to IAU can fail. Moreover, we assume that all authors are identical and rational, which also does not accord with reality. Thus, future work can further explore this extension of the model. Given that one can IAU at any point in the project and will still have to continue to work with the other authors on the project, determining the strategy of the optimal time to IAU would be of interest. Identifying this point might suggest what terms the authors need to agree on before the beginning of the project in order to eliminate such tactics in advance. In addition, future work can evaluate the optimal policy for the case of multiple research projects where there is a network of researchers and each research project is conducted by a different (or the same) subset of researchers. In such a scenario, a local improvement in an author's utility by IAU can produce more harm than good for the author in the long run. When his/her colleagues hear about the ultimatum, they may be less likely to collaborate with the author in the future. In particular, the effect of the network's size and density on the optimal policy would shed light on possible scenarios in which authors can exploit the ability to IAU without being "punished" by the community. Such information would provide the academic community with a data-driven tool to explore modifications to the current system in order to obtain the desired outcome.

\section*{Conclusion}
In an increasingly competitive academic landscape where co-authorship is gaining popularity as a tool to produce better, more sophisticated, and more comparative research, the current predominant academic system of co-authorship exposes authors to devastating outcomes in the form of a one-sided ultimatum from their peers. In order to better understand these dynamics, we develop a novel game-theory-based model for co-authorship disputes. Our model indicates that even for two authors, the average rate of issuing an ultimatum is 21\% and jumps to 43\% for five authors. Moreover, ironically, it is the authors who contributed the least who benefit the most from making ultimatums. 

Fortunately, it seems that the proposed model's predictions are more drastic than the real numbers, which are bounded by the cultural, personal, political, and sociological characteristics of the authors. Nonetheless, from a system's point of view, the current co-authorship paradigm indirectly encourages authors to engage in undesirable behavior. As such, the academic community should embrace a mechanism for co-authorship that penalizes authors for issuing ultimatums or obviates their ability to do so. While our model does not provide a solution, researchers in future works can utilize it as a benchmark for testing possible solutions. In addition, an extension of this model that includes the political, field-specific, and cultural characteristics of the authors is needed to accurately describe the complexity of the dynamics of co-authorship.

\matmethods{
Let us assume a fixed size set of authors, \(P\) \((N := |P|)\), participating in an academic co-authorship project. All the authors share the objective of finishing the project and publishing an academic manuscript. The utility of accomplishing this objective for each author \(p_i \in P\) can be different and denoted by \(u_i^{0}\) for the \(i_{th}\) author. In addition, each author has a utility distribution of being named in some position on the authors' list, denoted by \(u_{i}^{1}(j)\) for the \(i_{th}\) author at position \(1 \leq j \leq N\) on the authors' list. By default, the order of the authors' list is corresponds to contributions of each author. Thus, the overall utility of the \(i_{th}\) author of finishing the project and publishing the manuscript in position \(j\) on the authors' list is \(u_i^{0} \cdot u_i^{1}(j)\). 

We define a game with rounds \(t \in [1, \dots T]\) where \(T < \infty\) such that at \(t = T\) the project is completed and the manuscript is submitted for publication in an academic journal. The shared objective is accomplished once the authors have  contributed to a pre-defined threshold (\(\tau \in \mathbb{R}\)), which, therefore, directly defines \(T\). In each round, \(1 \leq t < T\), each author contributes in a stochastic manner by engaging in a wide range of actions such as conducting experiments, analyzing the results, and writing the manuscript.
Formally, the \(i_{th}\) author's contribution is denoted by \(\beta_i\), where \(\sum _{t=1}^{T} \sum _{i=1}^{N} \beta_i = 1\), \(\beta_i = f(\alpha_i)\), and \(f'(\alpha_i)<0\). In our model, we abstract the actions the author performs, focusing only on their abstract contribution. The utility distribution from the author's position on the authors' list has an impact on both his/her utility and the extent of his/her contribution to the project, denoted by \(u^{1} _{i}( \alpha_{i})\) such that its first and second derivative are negative for each value of \(\alpha_{i}\) and \(\frac{1}{N}< \alpha_{i} < 1\). In addition, an author can try to improve his/her position on the authors' list by issuing an ultimatum. In our context, the ultimatum takes the form of changing one author's position on the authors' list from \(j\) to \(k\) such that \(k < j\). If the ultimatum is not met, the proposing author can threaten to block the other authors from submitting the manuscript. In practice, this ultimatum is a real threat because all of the authors of a manuscript must approve its submission in order to publish it. Hence, if even one author is unwilling to provide his/her consent, none of the authors can achieve his/her shared objective. As such, all the authors, except the author who made the ultimatum, must accept the ultimatum. Formally, the process of responding to an ultimatum is as follows. First, each author (except for the author that who made the ultimatum) computes his/her own utility from accepting or rejecting the ultimatum and decides the better option for himself/herself. Second, if all of the authors accept the ultimatum, the position of the author who issued the ultimatum is set to \(1 \leq k \leq N-1\), and all authors with a position \(k\) or less are pushed down one position on the authors' list. If at least one author rejects the ultimatum, it is rejected. Then, the author who issued the ultimatum computes the utility of withdrawing the ultimatum or maintaining his/her stance (hold). If the ultimatum is withdrawn, the author who made the ultimatum receives a penalty of \(\rho \in \mathbb{R}\) that is reflected when computing his/her position on the authors' list with his/her contribution. Otherwise, the project ends and all authors suffer a negative utility according to the contribution they invested up to this point. Of note, given that the utility \(u^{0}\) is binary, at each point in time we assume that the \(i_{th}\) author considered a discounted utility \(\frac{u_i}{(1+\rho)^{T-t}}\), where \(\rho \in [0, 1]\) is the discount factor. Hence, an author would issue an ultimatum if it optimized his/her overall utility as indicated by:
\begin{equation}
    \max_{\alpha_i, \beta_i} u(\alpha_i, \beta_i) := \max_{\alpha_i, \beta_i} \frac{u^{0}_i}{(1+\rho)^{T-t}}+u^{1}_i(\alpha_i)+z_i(\beta_i)
    \label{eq:2}
\end{equation}
where \(z_i'(\beta_i)<0\) and \(z_i''(\beta_i)<0\). A schematic view of the ultimatum game and all its possible outcomes is provided in Fig. \ref{fig:ultimatum_game_view}. 

\begin{figure}[!ht]
    \centering
    \includegraphics[width=0.4\textwidth]{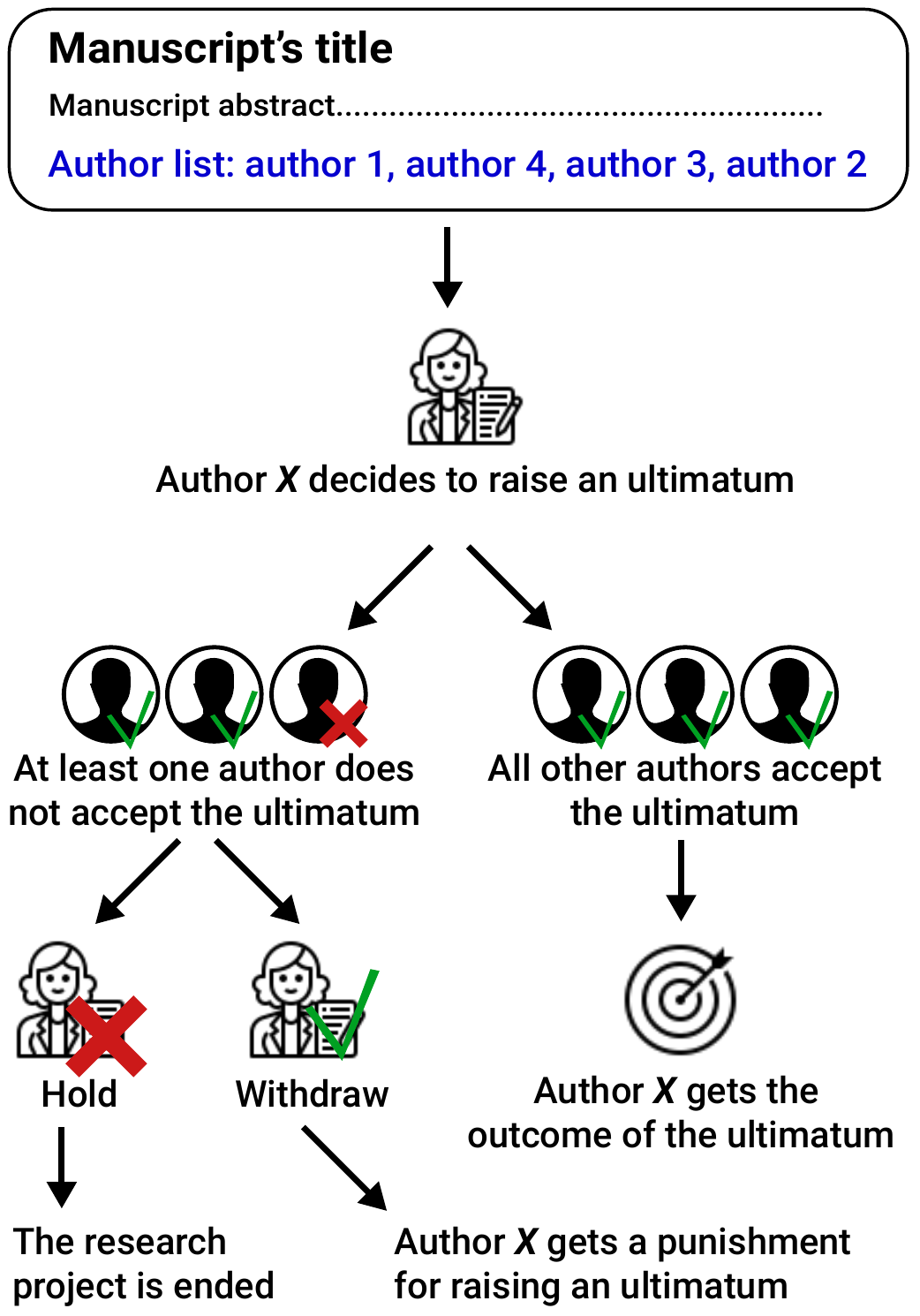}
    \caption{A schematic view of the process and possible outcomes from raising an ultimatum by an author.}
    \label{fig:ultimatum_game_view}
\end{figure}

\subsection*{Numerical simulation}
We implemented the proposed model as a computer simulation using the agent-based simulation approach \cite{agent_based_1,agent_based_2,agent_based_3}. Formally, each author (\(p \in P\)) is represented by a finite state machine, defined by the tuple \(p := \{c, w_m, w_s, u^{0}, u^{1}, \rho\}\) where \(c \in \mathbb{R}\) is the current amount of contribution the author have provided toward the shared objective, \(w_m, w_s \in \mathbb{R}\) are the mean and standard deviation of the author's contribution to the shared objective in every single step in time, \(u^{0} \in \mathbb{R}\) is the personal utility from accomplishing the shared objective (i.e., publish the manuscript), and \(u^{1} \in \mathbb{R}^{|P|}\) is the personal utility from being in each position in the authors' list. 

At each round, in a random order, the authors contribute to the shared objective by sampling from the normal distribution, \(N(w_{m}, w_{s})\). In addition, each author can issue an ultimatum. In the case which an author made an ultimatum , it is resolved immediately after the author contributed to the shared objective. As part of issuing an ultimatum, an author decides which position \(i\) to ask for, given his/her current position \(j\). An author would issued an ultimatum for position \(i\) given current position \(j < i\) if and only if the following statement is fulfilled:
\begin{equation}
    \forall p \in P: c > u^{0} ( u^{1}(m) - u^{1}(l), T-t),
    \label{eq:offer_chicken}
\end{equation}
where \(m\) is the current position of an author on the authors' list and \(l\) is its new position. For example, if \(i < m < j\) so \(l = m - 1\).

For all the simulations, the personal utility function \(u^{1}\) of the authors take the form \(u^{1}(x) = \big ( 1-R(0, 0.25) \big ) / \big ( x+R(0, 0.25) \big )\) where \(x\) is the position of the author on the authors' list and \(R(a,b)\) is a function that receives two positive numbers \(a\) and \(b\), such that \(a < b\) and returns a random number \(a \leq b \leq c\) in a evenly distributed manner. If not stated otherwise, the default model parameters used in the simulations are shown in Table \ref{table:parameters}.

\begin{table}[!ht]
\centering
\begin{tabular}{|l|c|c|}
\hline
\textbf{Parameter} & \textbf{Value} \\ \hline
Authors' utility from finishing the project spectrum [1] & \(R(1,5)\) \\ \hline
Authors' contribution spectrum [1] & \(R(1,5)\) \\ \hline
Project duration in weeks (\(T\)) [1] & \(R(8, 88)\) \\ \hline
Point in the project percent [1] & \(R(0, 100)\) \\ \hline
Number of authors (\(N\)) [1] & \(R(2, 8)\) \\ \hline
\end{tabular}
\caption{Default model's parameters' values. }
\label{table:parameters}
\end{table}

The implementation of the proposed numerical simulation and the model is freely available on the project's GitHub page\footnote{The simulation's implementation is freely available at \url{https://github.com/teddy4445/coauthorship_simulator}}.
}

\showmatmethods{} 

\showacknow{} 

\bibliography{pnas-sample}

\end{document}